# Encoding and multiplexing information signals in magnetic multilayers with fractional skyrmion tubes

*Runze Chen[1], Yu Li[1], Will Griggs[1], Yuzhe Zang[1], Vasilis F. Pavlidis[2], and Christoforos Moutafis[1]*

[1]Nano Engineering and Spintronic Technologies (NEST) research group, Department of Computer Science, The University of Manchester, Manchester M13 9PL, United Kingdom

[2]Advanced Processor Technologies (APT) research group, Department of Computer Science, The University of Manchester, Manchester M13 9PL, United Kingdom

**KEYWORDS** Skyrmions, Spintronics, Skyrmionics, Magnetic multilayers, Micromagnetics, Multiplexer.

**ABSTRACT**

Tailored magnetic multilayers (MMLs) provide skyrmions with enhanced thermal stability, leading to the possibility of skyrmion-based devices for room temperature applications. At the same time, the search for additional stable topological spin textures has been under intense research focus. Besides their fundamental importance, such textures may expand the information encoding capability of spintronic devices. However, fractional spin texture states within MMLs in the vertical dimension have yet to be investigated. In this work, we demonstrate numerically fractional skyrmion tubes (FSTs) in a tailored MML system. We subsequently propose to encode sequences of information signals with fractional skyrmion tubes (FSTs) as information bits in a tailored MML device. Micromagnetic simulations and theoretical calculations are used to verify the feasibility of hosting distinct FST states within a single device, and their thermal stability is investigated. A multilayer multiplexing device is proposed, where multiple sequences of the information signals can be encoded and transmitted based on the nucleation and propagation of packets of FSTs. Finally, pipelined information transmission and automatic demultiplexing is demonstrated by exploiting the skyrmion Hall effect and introducing voltage-controlled synchronizers and width-based track selectors. The findings indicate that FSTs can be potential candidates as information carriers for future spintronic applications.

**INTRODUCTION**

Magnetic skyrmions are particle-like topological spin configurations [1,2]. They are promising candidates as information carriers in future information technologies, owing to their topologically protected morphological stability and dynamical properties [2,3]. This stability comes from the competition between magnetic interactions which favor colinear spin configurations, such as Heisenberg exchange coupling and perpendicular magnetic anisotropy (PMA), and interactions favoring orthogonal configurations such as dipolar coupling and Dzyaloshinskii-Moriya interaction (DMI) [4,5]. In both bulk magnetic and magnetic multilayer (MML) systems with broken symmetry, skyrmion spin states become one of the system energy minima. The magnetic skyrmion can be described by an integer topological index, called the skyrmion number $N$, which counts how many times the magnetisation wraps around a unit sphere. The skyrmion number is defined as [6]:

$$N = \frac{1}{4\pi} \int \boldsymbol{m} \cdot \left(\frac{\partial \boldsymbol{m}}{\partial x} \times \frac{\partial \boldsymbol{m}}{\partial y}\right) dx dy \qquad (1)$$

where $\boldsymbol{m}$ is the normalized magnetisation and $N = \pm 1$ corresponds to the case of magnetic skyrmions, with the sign reflecting the polarity. Skyrmions can be competitive candidates as information carriers in low power and highly efficient computational devices because of their non-volatility, nanoscale size, and ease of manipulation [3]. These advantages have inspired proposals for their implementation in skyrmionic transistors [7], logic gates [8–10], racetrack memory [11–13], nano-oscillators [14], resonant diodes [15], neuromorphic computing [16,17], and reservoir computing [18]. Notably, Ref. [17] is an experimental demonstration of a skyrmionic synapse device, while the other references contain numerical results.

For each of these applications, room temperature operation is a critical requirement for realistic device integration. Recently, tailored MMLs have been explored as a means to host skyrmions at room temperature [19,20], where the stacking of a repeated layer structure leads to enhanced thermal stability. This stability in MMLs can be attributed to the increased DMI from the asymmetric interfaces [Heavy metal ($HM_1$)|Ferromagnetic (FM)] and [FM|$HM_2$] and the increased magnetic volume of skyrmions. Additional to the requirement for room-temperature stability, in order for skyrmions to be practically used in functional devices it is highly desirable for them to be stabilized under zero applied magnetic field. Recent work has highlighted this possibility, for example by leveraging lateral confinement effects [21] or by careful engineering of interlayer exchange coupling [22].

At the same time, there has been much recent effort to find skyrmion-like quasiparticles with topological charge other than ±1. Such skyrmionic quasiparticles have been proposed as information carriers to enhance device functionality [23,24]. For example, skyrmionium, which comprises a skyrmion-antiskyrmion pair, is a type of magnetic quasiparticle with a vanishing topological charge $N = 0$ which has been proposed for use in racetrack memory applications [23]. Similarly, anti-skyrmionite (i.e., a double-antiskyrmion-skyrmion pair with $N = \mp 1$ [25]) has the potential to be used as an additional information carrier in skyrmionic devices. More recently, several studies have shown that nanomagnets can indeed host a plethora of topological and non-topological quasiparticles, including both theoretical calculations [24] and experimental demonstrations of skyrmion bags in liquid crystals [26] and skyrmion bundles in chiral magnets [27]. However, these proposals are explorations of skyrmionic particles limited in two-dimensional systems. A natural next step is to explore skyrmion-like quasiparticles which are distinguishable by their three-dimensional profile, including variations in the film-perpendicular direction. In fact, partially stabilised skyrmion quasiparticles have recently been theoretically predicted in bulk chiral magnets, including chiral bobbers [28] and dipole strings [29], enriching the diversity of the skyrmionic family.

Previous studies in MML systems have mainly focused on skyrmion 'tube' states, in which skyrmions are stabilised throughout the MML stack. Standard tubular skyrmions are effectively stacks of skyrmions in neighbouring magnetic layers which exist across the whole multilayer structure [20]. There has been much recent interest in exploring topological magnetic textures in 3 dimensions by engineering the properties of heterostructures so as to host complex depth-dependent states [30]. Our proposal is to use fractional skyrmion tubes (FSTs) in MMLs to encode multiple information signals. These FSTs comprise coupled skyrmions that occupy discrete fractions of the full multilayer system. This approach enables us to investigate skyrmion tubes with distinct depth profile, making it possible to explore multiple states within a single system. We will show that such fractional skyrmion tubes can exhibit effective tunability in their thermal stability. Based on their topological properties and propagation by electric currents, we propose a multilayer multiplexing device which relies on the nucleation, propagation, and automatic selection of multiple distinct FSTs in a single device. Our proposal in this work highlights the potential of encoding information via distinct FST states in MMLs. Specifically, we use simulations to demonstrate pipelined information transmission and automatic demultiplexing of information signals via use of voltage-controlled synchronizers and track selectors, respectively. Most results of this work are performed via the micromagnetic simulation package Mumax$^3$ [31]. Mathematical calculations are performed via the Python mathematical module SciPy [32].

RESULTS

1. Fractional skyrmion tubes in MMLs as multi-bit information carriers

1.1. Potential multi-bit information carriers

As aforementioned, tailored MMLs have been proposed to stabilize room temperature skyrmions, typically as skyrmion tubes which extend vertically throughout the depth of the MML [33,34]. In contrast, here we investigate the possibility of obtaining fractions of full skyrmion tubes, wherein skyrmions are stabilised in select layers of the system. All FST states studied in this work comprise Néel-type skyrmions in individual layers, which are favoured over Bloch-type skyrmions in multilayer structures with interfacial DMI. Fig. 1(a) shows an example of a 4MML nanotrack hosting four distinct skyrmion states, each comprising a stack of skyrmions which extend over one, two, three, or all four layers of the system. Such skyrmion states are hereafter referred to as fractional skyrmion tubes (FST). Thus, Fig. 1(a) exhibits four FSTs, namely (from left to right) a 4MML FST, a 1MML FST, a 3MML FST, and a 2MML FST. We will show that each of these can be nucleated and manipulated individually, leading to a fourfold enhancement of the capacity of the 4MML device to store and process information.

To demonstrate the effect of the micromagnetic parameters on the stability of the four distinct FST states in our example 4MML system, we conducted micromagnetic simulations which explore the phase diagram of FSTs by modifying i) the external out-of-plane magnetic field, ii) the DMI constant, and iii) the ferromagnetic interlayer exchange coupling. The interlayer exchange coupling modelled throughout this work is of RKKY type and is always positive such to favor parallel alignment between spins in neighbouring layers. In physical multilayered systems, the magnitude and sign of the interlayer exchange coupling has an oscillatory dependence on the magnetic layer separation, with the exact character of this sensitivity depending on the particular material structure. Therefore, the interlayer exchange coupling can be effectively modified by tuning the thickness or changing the spacer layer material [35–37], where either ferromagnetic or antiferromagnetic interlayer coupling can be achieved. In this work, we model changes to the interlayer exchange coupling without consideration for any particular material system, and therefore do not model the corresponding changes to material thickness. To model the RKKY-type interlayer exchange interaction in the micromagnetic simulations, we implement a custom field into the Mumax3 code which couples nearest neighbouring magnetic layers. Next-nearest neighbour and higher order interactions between the magnetic layers are assumed to be negligible. The phase diagrams of distinct FST states in a 4MML film with $J_{\text{Interlayer}} = 0$ and $J_{\text{Interlayer}} = 0.04$ mJ m$^{-2}$ is shown in Fig. 1(b).

In the simulations, the external magnetic field was scanned in the range of 10 to 100 mT with a 10 mT step applied out-of-plane, and the DMI constants were varied from 1.5 to 2.6 mJ m$^{-2}$ with a step of 0.1 mJ m$^{-2}$, while keeping the interlayer exchange coupling constant at 0 and 0.04 mJ m$^{-2}$. For each data point in Fig. 1(b), we configured the system with an initial ansatz that contained a 1MML FST, a 2MML FST, a 3MML FST, and a 4MML FST, respectively. We then allowed the system to equilibrate with the corresponding initial states and magnetic parameters. Each cell of Fig. 1(b) shows the number of distinct FST states which are stabilised in the equilibrated system. For instance, a "0" signifies that we obtain no FST states (i.e. we have the ferromagnetic (FM) state), while a "4" denotes that all of the 1MML FST, 2MML FST, 3MML FST, and 4MML FST states can be stabilised and distinctly identified. Note that we only record the total number of distinct states stabilised in each case; we do not distinguish between different combinations of FSTs.

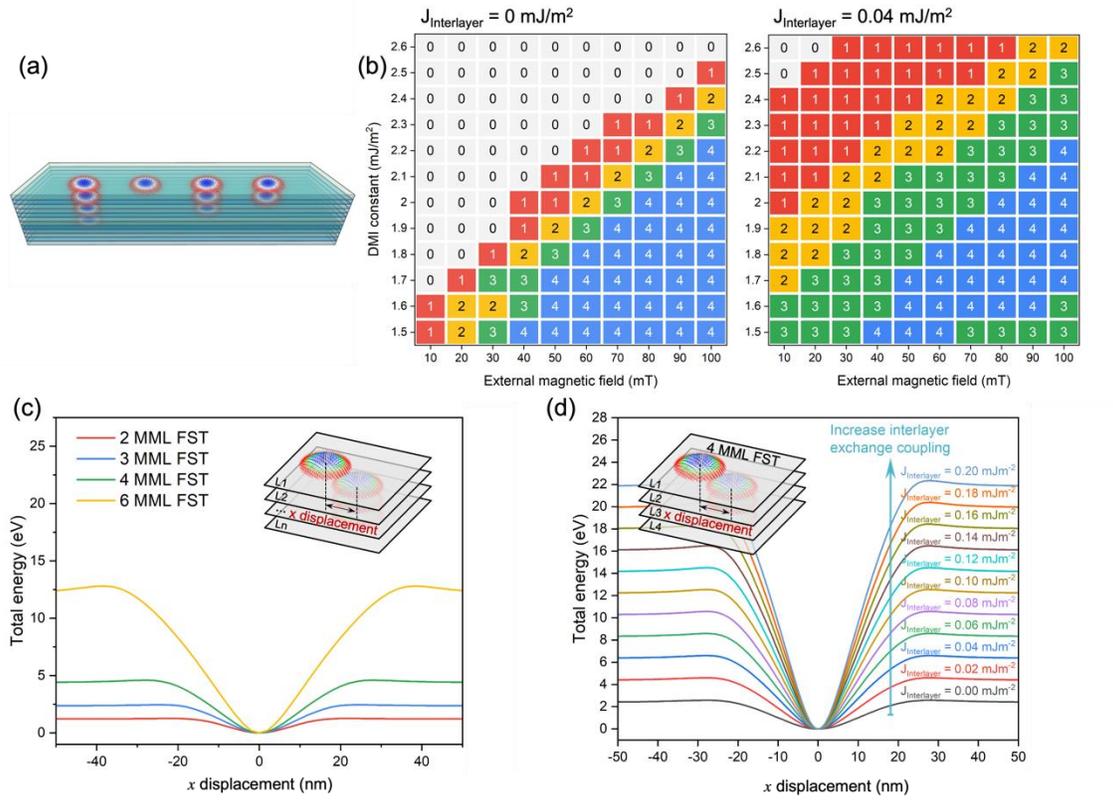

**Figure 1.** Fractional skyrmion tubes in MMLs. (a) Schematic illustration of the 4MML, 1MML, 3MML, and 2MML FSTs, respectively, stabilised in a 4MML system. (b) Phase diagram of the number of distinct FST states with varied external out-of-plane magnetic field and DMI constant in a 4MML film without and with the interlayer coupling. The integers 0-4 represent the number of distinct states. (c) Energy change of the 2MML, 3MML, 4MML, and 6MML FSTs as a function of the $x$ displacement of the top layer skyrmion with the interlayer exchange constant $J_{\text{Interlayer}}$ = 0.02 mJ m$^{-2}$. (d) Energy change of the 4MML FST as a function of the $x$ displacement of the top layer skyrmion with the interlayer exchange coupling constant ranging from 0 to 0.20 mJ m$^{-2}$. Magnetic parameters in simulations of (c) and (d): external magnetic field 90 mT and DMI constant 1.9 mJ m$^{-2}$ which are selected from (b) in order to obtain stable FSTs states in the system.

According to the results shown in Fig. 1(b), there is a noticeable transition from the no FST state at a low out-of-plane magnetic field and high DMI constant to four distinct FST states at a high field and low DMI constant. There is a broad parameter window for stabilizing four distinct FST states in the 4MML systems without interlayer exchange coupling. However, with increasing the interlayer exchange coupling, the parameter window for four FST states shrinks, while the parameter windows for three FST states and one/two FST states expand towards the initial 4 FST phase and 0 FST phase, respectively (see also the phase diagram at higher interlayer exchange coupling constant $J_{Interlayer}$ = 0.08 to 0.12 mJ m$^{-2}$ in Fig. S1 of the Supporting Information). We note that even small values of the interlayer exchange coupling are sufficient to substantially alter the stability of FSTs. We can understand these changes by considering the roles of DMI and external magnetic field, which tend to favour perpendicular and parallel spin configurations respectively. At a given value of the external field, if the DMI is too large then the labyrinth domain state becomes favourable; if it is too small then colinear spins are favoured. It follows that the effect of the interlayer exchange, which favours parallel spins, is to reduce the required external magnetic field and increase the allowable DMI for FSTs to be stabilised. Hence, the interlayer exchange coupling has the effect of widening the parameter windows for stabilizing 1MML, 2MML, and 3MML FSTs. We additionally observe that for sufficiently large external field and sufficiently small DMI (i.e. the bottom right corner of the phase diagram) the stability of FSTs is diminished by the presence of interlayer exchange coupling. Here, the combined effect of the interlayer exchange and large external field means that larger values of the DMI are required in order to stabilize chiral spin textures such as FSTs. Therefore, by modifying the magnetic parameters of the material system, we can effectively tune the distinct states phase diagram to achieve the maximum number of distinct FST states.

Apart from the results exhibited in Fig. 1(b), we also performed simulations on an 8MML film to demonstrate the extension of the FSTs in even higher-level cascaded MMLs. As shown in Fig. S2 of Supporting Information, we can obtain eight distinct FST states with a large parameter window in the 8MML film. A similar transition to that seen in Fig. 1(b) is also observed. Similarly, by tunning the interlayer ferromagnetic exchange coupling $J_{\text{Interlayer}}$ from 0 to 0.08 mJ m$^{-2}$, we observe an apparent shrinking of region corresponding to the larger number of distinct FST states. In summary, the data in Fig. 1(b) shows that to maximize the number of distinct FST states, a relatively high external magnetic field (70 to 100 mT), a moderate DMI constant (1.8 to 2.2 mJ m$^{-2}$), and a relatively weak interlayer ferromagnetic exchange coupling (lower than 0.08 mJ m$^{-2}$) are required.

### 1.2. Thermal stability of fractional skyrmion tubes

To reveal the reliability and feasibility of using multiple FSTs in a single MML device, it is important to assess their thermal stability. There exist multiple numerical approaches for characterising the thermal stability of magnetic textures. The most widely used is the geodesic nudged elastic band method (GNEBM) [38] which can identify the internal energy difference between an initial and a final spin texture. In the present work, we provide a simple analysis of the thermal stability based on the Arrhenius-Néel law by using the difference in internal energy between the intact and decoupled FST states to be their binding energy. By calculating the binding energy, it is possible to characterise the thermal stability of FST states at finite temperatures. We thus determined the binding energy of different FSTs as a function of the in-plane separation between two nearest-neighbour skyrmions in the two topmost layers of the FST stack, respectively. The binding energy of FSTs was calculated according to Ref. [39]; the results are shown in Fig. 1(c). First, we relaxed the skyrmions in a 2MML FST, 3MML FST, 4MML

FST, and 6MML FST, respectively. Then we artificially shifted the top-layer skyrmion from -50 nm to 50 nm in the *x*-direction while fixing the position of skyrmions in the other layers. The total micromagnetic energy of the whole system at every *x*-shifted position is calculated without relaxing and equilibrating the system. As shown in Fig. 1(c), we normalised the results by subtracting the energy of initial FST states from the energy after shifting the top layer skyrmion. Therefore, the binding energy as a function of *x*-shifted position in Fig. 1(c) is greater or equal to 0. The relative value represents the increased total energy induced from shifting the top layer skyrmion.

The binding energy in Fig. 1(c) shows that by shifting the top layer skyrmion away from the centre, the total energy first rises gradually and then remain approximately constant after an *x*-displacement of ~35 nm. The difference between the highest energy state and the initial state is the binding energy of the FST, which quantifies the energy barrier separating the FST and decoupled skyrmion states. The interlayer exchange coupling $J_{\text{Interlayer}}$ = 0.02 mJ m$^{-2}$, external magnetic field of 90 mT, and DMI constant of 1.9 mJ m$^{-2}$ were used when simulating FSTs in Fig. 1(c). The binding energy $E_b$ of the 2MML FST, 3MML FST, 4MML FST and 6MML FST are calculated as 1.26 eV, 2.46 eV, 4.60 eV, 12.81 eV, respectively. The increased binding energy from 2MML FST to 6MML FST can be attributed to enhanced dipolar coupling.

To demonstrate the tunability of the FST thermal stability, we next modified the interlayer exchange coupling constant $J_{\text{Interlayer}}$ from 0 to 0.08 mJ m$^{-2}$ for each FST. The binding energy of FSTs is calculated and displayed in Fig. S3 of the Supporting Information. It is observed that the binding energy of each FST follows a linear relationship with the interlayer exchange coupling of the MMLs. To explore this, we calculated the binding energy of a 4MML FST with $J_{\text{Interlayer}}$ ranging from 0 to 0.2 mJ m$^{-2}$. The results, shown in Fig. 1(d), verify that the binding energy increases with interlayer exchange coupling. Similar results can also be observed for 2MML, 3MML, and 6MML FSTs, as shown in Fig. S4 of the Supporting Information.

The larger calculated binding energy results in an improved thermal stability. The thermal stability can be quantified using the Arrhenius-Néel law to estimate the lifetime of a metastable state [38,40],

$$\tau(E_\mathrm{b}) = \tau_0 \exp\left(\frac{E_\mathrm{b}}{k_\mathrm{B} T}\right), \tag{2}$$

where $f = \tau_0^{-1}$ is attempt frequency, $k_\mathrm{B}$ is the Boltzmann constant, $T$ is the temperature under consideration, and $E_\mathrm{b}$ is the binding energy, as given by the analysis above. Here, we assume $T = 300$ K in order to estimate the FST lifetime at room temperature. An attempt frequency of $10^9$ - $10^{12}$ Hz is typically used [38,40,41]. However, there is some debate about the correct value of the attempt frequency to use [42], which can be as large as $10^{21}$ Hz. Precisely estimating the lifetime of quasiparticles is beyond the scope of this paper. We therefore choose a large value of $10^{21}$ Hz for the attempt frequency here and give a conservative estimation of the lifetime of FSTs. The binding energy of FSTs extracted from Fig. 1(c) with $J_\mathrm{Interlayer} = 0.02$ mJ m$^{-2}$ lead to estimated lifetimes of 20.1 ns, 1.47 s, 2.12×10$^{20}$ s, 1.89×10$^{56}$ s, and 1.58×10$^{194}$ s for the 1MML FST, 2MML FST, 3MML FST, 4MML FST, and 6MML FST, respectively. Note that the estimated results merely reflect the thermal stability and annihilation probability, rather than the precise lifetime of FSTs in realistic devices. As we are using multiple FSTs in a single device, the least stable one defines the thermal stability of the whole device. However, we can enhance the lifetime of each FST by 5 orders of magnitude by tuning the magnetic parameters and interlayer exchange coupling. Our results here demonstrate the tunability of the thermal stability of FSTs, a key consideration for experimental and commercial device design.

## 2. Magnetic and topological properties of fractional skyrmion tubes

To utilize FSTs in realistic devices and applications, it is necessary to evaluate their magnetic and topological properties, especially the propagation behaviour under the applied electric current. Fig. 1 demonstrates that the various MML FSTs have different dimensions in the *x-y* plane. We thus stabilised a series of FSTs in an 8MML film to demonstrate the MML-dependent diameter of skyrmions.

In the simulations, we artificially equilibrated skyrmions in the select MMLs in order to obtain a 1MML FST, 2MML FST, 3MML FST, 4MML FST, 5MML FST, 6MML FST, 7MML FST, and 8MML FST. The layers are marked as L1 to L8 for the 8MML film in Fig. 2(a), where the bottom layer is L1, and the top layer is L8, such that the stabilised 1MML FST contains only one skyrmion in L1 while L2 to L8 remain in the saturated FM state. Similarly, the 4MML FST has skyrmions within L1 to L4, and 8MML FST has skyrmions within all layers as a full tube. The relative size and skyrmion diameter of these FSTs can be visualized by superposing 1MML to 8MML FSTs together, as shown in Fig. 2(a). The skyrmion diameter of FSTs grows monotonically with the number of MMLs. This trend can be attributed to the increased dipolar field as the number of cascaded skyrmions increases [16]. The measured skyrmion diameter of FSTs is shown by the black lines in Figs. 2(c) and 2(d), where an approximately linear relation with the number of MMLs can be observed. The fact that the diameter increases with the number of layers occupied by the FST is likely a result of the enhanced dipolar coupling as the vertical extend of the FST increases.

We then investigated the transport behaviour of different MML FSTs under electric current. The spin-orbit torque (SOT) induced by a current perpendicular to the plane (CPP) geometry and the spin-transfer torque (STT) from a current in plane (CIP) geometry are considered. For the CIP case, the charge current is applied in the FM layer in the *x*-direction with a density of 15 MA cm$^{-2}$ within each layer of the MML. In the case of CPP, a charge current with density 50 MA cm$^{-2}$ is applied in the bottom HM layer only; the thickness of the HM layers in each MML is smaller than the electron diffusion length [43] so that no spin Hall effect will be induced [44]. A spin current with spin polarisation in +*y* direction is created and injected into the first MML starting from the bottom, namely L1 mentioned above. As a result, FSTs move along a trajectory at an angle to the direction of the applied current, which is the well-known skyrmion Hall effect (SkHE) [45]. The angle between the FST trajectory and the direction of applied current (+*x* in this example) is defined as the skyrmion Hall angle $\theta_{SkHE}$. Under the SOT, FSTs propagate along the +*y* transverse direction, while under the STT, FSTs move towards the opposite transverse direction. As shown in Fig. 2(b), the 1MML FST has the largest $\theta_{SkHE}$, while the 4MML FST exhibits a smallest $\theta_{SkHE}$ for both CPP and CIP.

To better understand the simulated spin dynamics, it is instructive to employ an analytical model for the motion of non-collinear spin textures. Therefore, we utilize the Thiele equation by imposing the stationary limit that the FSTs move with a constant velocity and that the texture does not deform. We consider both CIP and CPP geometries using the Zhang-Li [46] and Slonczewski [47] torques in the LLG equation (see Methods), respectively.

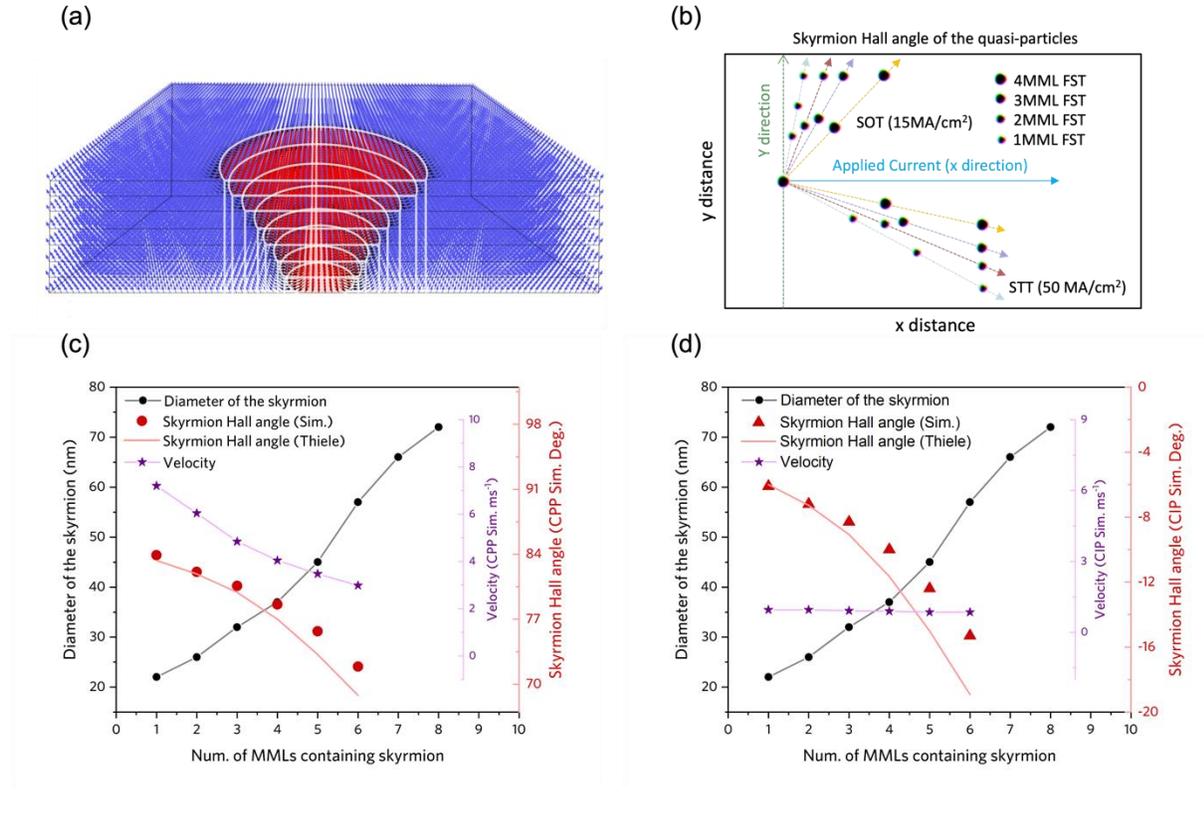

**Figure 2.** Magnetic and topological properties of fractional skyrmion tubes. (a) Cross-sectional view of the 1MML to 8MML FSTs stabilised in the same 8MML system. White outlines of the FSTs indicate the varying skyrmion diameter from 1MML to 8MML. (b) Simulated trajectories of FSTs driven by SOT with the current density of 15 MA cm$^{-2}$ and a spin Hall ratio of 0.6, and STT with the current density of 50 MA cm$^{-2}$. (c) and (d) show simulation results and theoretical predictions from the Thiele equation of the skyrmion diameter, the velocity of movement, and skyrmion Hall angle for 1MML to 8MML FSTs under (c) SOT with current perpendicular to the plane and (d) STT with the current in the plane. Note that 7MML and 8MML FSTs transformed to the stripe domain wall states after applying the electric current. The magnetic parameters of simulations are interlayer exchange coupling constant $J_{Interlayer} = 0.02$ mJ m$^{-2}$, external magnetic field 90 mT, and DMI constant 1.9 mJ m$^{-2}$.

In the CPP geometry, assuming a periodical boundary condition in the x-y plane, the translational motion of spin textures driven by the spin Hall effect can be described by a modified Thiele's equation [48,49]

$$\boldsymbol{G} \times \boldsymbol{v} - \alpha \boldsymbol{\mathcal{D}} \cdot \boldsymbol{v} - \mathcal{T}_{SOT}\boldsymbol{\mathcal{J}} \cdot \mathbf{m}_\text{p} = 0, \quad (3)$$

where $\boldsymbol{G} = (0,\ 0, -4\pi N)$ is the Gyroscopic vector with the topological charge $N$ defined in Eq. 1, and $\boldsymbol{v} = (v_\text{x},\ v_\text{y})$ is the skyrmion drift velocity along the $x$ and $y$ axes, respectively. The first term $\boldsymbol{G} \times \boldsymbol{v}$ in Eq. 3 is the topological Magnus force that results in the transverse motion of skyrmions as a function of the driving current, which directly results in the SkHE [45]. $\alpha$ is the magnetic damping parameter, and $\boldsymbol{\mathcal{D}}$ is the dissipative tensor which is calculated by $\mathcal{D}_{ij} = \frac{1}{M_s^2}\iint \frac{\partial \boldsymbol{M}}{\partial x_i} \cdot \frac{\partial \boldsymbol{M}}{\partial x_j} dxdy = \begin{bmatrix} \mathcal{D}_{xx} & \mathcal{D}_{xy} \\ \mathcal{D}_{yx} & \mathcal{D}_{yy} \end{bmatrix}$. The term $\mathcal{T}_{SOT}\boldsymbol{\mathcal{J}} \cdot \mathbf{m}_\text{p}$ quantifies the effect of the SOT over the magnetic quasiparticle, where $\mathcal{T}_{SOT} = \frac{\gamma_e \hbar j_e \theta_{SH}}{2eM_s t}$ is the amplitude of SOT over the quasiparticle, $\gamma_e = \frac{\gamma}{\mu_0} = 1.76 \times 10^{11}\ T^{-1}\ s^{-1}$ is the gyromagnetic ratio of an electron, $\hbar$ is the reduced Planck constant, $j_e$ is the current density, $\theta_{SH}$ is the spin Hall ratio, $e$ is the electron charge, $M_s$ is the saturation magnetisation, and $t$ is the thickness of the FM layer. $\boldsymbol{\mathcal{J}}$ is the driving torque tensor which is calculated by $\mathcal{J}_{ij} = \frac{1}{M_s^2}\iint \left(\frac{\partial \boldsymbol{M}}{\partial x_i} \times \boldsymbol{M}\right)_j dxdy = \begin{bmatrix} \mathcal{J}_{xx} & \mathcal{J}_{xy} \\ \mathcal{J}_{yx} & \mathcal{J}_{yy} \end{bmatrix}$, and $\mathbf{m}_\text{p}$ is the polarisation direction of the spin current. By solving Eq. 3, we can obtain the skyrmion Hall angle of the FSTs as

$$\theta_{\text{SkHE(CPP)}} = \arctan\left(\frac{v_y}{v_x}\right) = -\frac{4\pi N}{\alpha \mathcal{D}_{xx}}. \quad (4)$$

In CIP geometry, the Thiele equation has the form [11,49]

$$\boldsymbol{G} \times (\boldsymbol{j} - \boldsymbol{v}) + \boldsymbol{\mathcal{D}}(\beta\boldsymbol{j} - \alpha\boldsymbol{v}) = 0, \tag{5}$$

where $\boldsymbol{G}$, $\boldsymbol{\mathcal{D}}$, $\alpha$, and $\boldsymbol{v}$ are as defined above. $\beta$ is the non-adiabaticity of spin-transfer-torque, for which a small value of 0.02 is used, and $\boldsymbol{j}$ is the current vector describing the direction and amplitude of the applied current. From Eq. 5, the skyrmion Hall angle of FSTs under CIP is therefore

$$\theta_{\text{SkHE(CIP)}} = \arctan\left(\frac{v_y}{v_x}\right) = \frac{4\pi(\alpha-\beta)\mathcal{D}_{xx}N}{\alpha\beta\mathcal{D}_{xx}^2 + 16\pi^2 N^2}. \tag{6}$$

We calculated $\theta_{\text{SkHE}}$ for both CPP and CIP by Eq. 4 and Eq. 6, respectively. Owing to the fact that different FST states have different diameters, the calculated value of $D_{xx}$ is different for each FST state. Hence, the skyrmion Hall angle varies with the number of magnetic layers occupied by the FST. Comparison of the theoretical prediction and simulation results of $\theta_{\text{SkHE}}$ for 1MML to 8MML FSTs are exhibited in Figs. 2(c) and 2(d), where Fig. 2(c) is for CPP geometry and Fig. 2(d) is for CIP. The simulated results of $\theta_{\text{SkHE}}$ are shown by the discrete data points, and the calculated results from Thiele's equation are exhibited as solid lines. There is a decrease of $\theta_{\text{SkHE}}$ both in simulations and theoretical calculations when cascading MMLs for FSTs in both CPP and CIP geometry. However, there is a slight disparity between the theoretical predictions and simulations for $\theta_{\text{SkHE}}$, which becomes more significant as the number of MML repeats increases. The overestimation of skyrmion Hall angle by the Thiele equation can be explained by additional dissipation mechanisms related to dynamic variation of the skyrmion shape, which cannot be captured by the rigid shape approximation used to derive the Thiele equation.

We also extracted the velocity of FST propagation in CPP and CIP from simulations. The results are shown in Figs. 2(c) and 2(d). FSTs generally moves faster in CPP than in CIP, even with a smaller amplitude of the applied electric current. This trend is consistent with previous studies comparing CPP and CIP motion, and likely results from the more significant Slonczewski in-plane torque in CPP geometry compared to the smaller, field-like torque which results from CIP [2,13]. On the other hand, a marked drop of FST velocity from 7.5 ms$^{-1}$ for 1MML FST to 3 ms$^{-1}$ for 6MML FST under the same driving current can be observed in the CPP geometry, while the velocity of FSTs remains virtually constant as the number of MML repeats increases for the case of CIP geometry. The different current-dependent characteristics of skyrmion velocity could result from the way that torque is injected into the magnetic textures via the CPP and CIP geometry. In the CIP geometry, the torques act on individual skyrmions in each layer, while in the CPP geometry, the torque is merely injected from the top layer skyrmion so that velocity would decrease as the number of MML repeats increases. Overall, these results indicate that the proposed FSTs have tuneable thermal stability and distinct topological and magnetic properties, which may be used as signatures to identify them from each other. In the next chapter, we demonstrate one of the potential uses for FSTs: an MML multiplexing device.

## 3. Magnetic multilayer multiplexing device with fractional skyrmion tubes

### 3.1. Demonstration of the proposed multiplexing device

In this section, a multilayer multiplexing device is proposed using multiple FSTs as information carriers. Such a device can perform signal multiplexing, signal transmission, and automatic signal demultiplexing. We have shown that as many as eight distinct FST states, possibly more, can be achieved through careful selection of the system and magnetic parameters. In this section, an FST-based device built from a 4MML nanotrack is illustrated for simplicity. The schematic of the proposed 4MML multiplexing device is illustrated in Fig. 3. The multiplexing device has three parts: a 4MML nanotrack for FST transmission, terrace-like MML stages for FST nucleation via magnetic tunnel junctions (MTJs) fabricated on the top, and a four-branched track for FST automatic selecting.

Experimental fabrication of this device will require a complex, multiple-step patterning process. For example, the terrace structure for FST nucleation may be fabricated via several subtractive lithography steps, with single layer removal achieved through carefully calibrated ion-beam milling or focused ion beam etching. Given the modular nature of the device operation, it is in principle possible to optimise each component of its functionality (i.e. the FST stability, nucleation, chambering, propagation, and detection) individually, with the full device operation verified thereafter. Indeed, some of the required steps have recently begun to be experimentally explored, including current-induced skyrmion nucleation in MTJs [50] and VCMA control of their transport through nanotracks [51].

The proposed device shown in Fig. 3 can multiplex and transmit four distinct sequences of information signals simultaneously, where each information signal is encoded by one type of FST, and the presence/absence of the FST encodes information "1"/"0". As shown in Fig. 3, each type of FST acts as a distinct information carrier. The workflow of the proposed multilayer multiplexing device contains four procedures: 1) FSTs are nucleated at the terrace-like MML stages via STT by injecting electric current from MTJs; 2) Multiple FSTs are chambered into the 4MML nanotrack in preparation for the transmission of signals; 3) Information transmission via FSTs propagation along the 4MML nanotrack; 4) Automatic demultiplexing of the information signals (FSTs) via the SkHE and the four-branched track selector. These four procedures are explored in detail in the following sections.

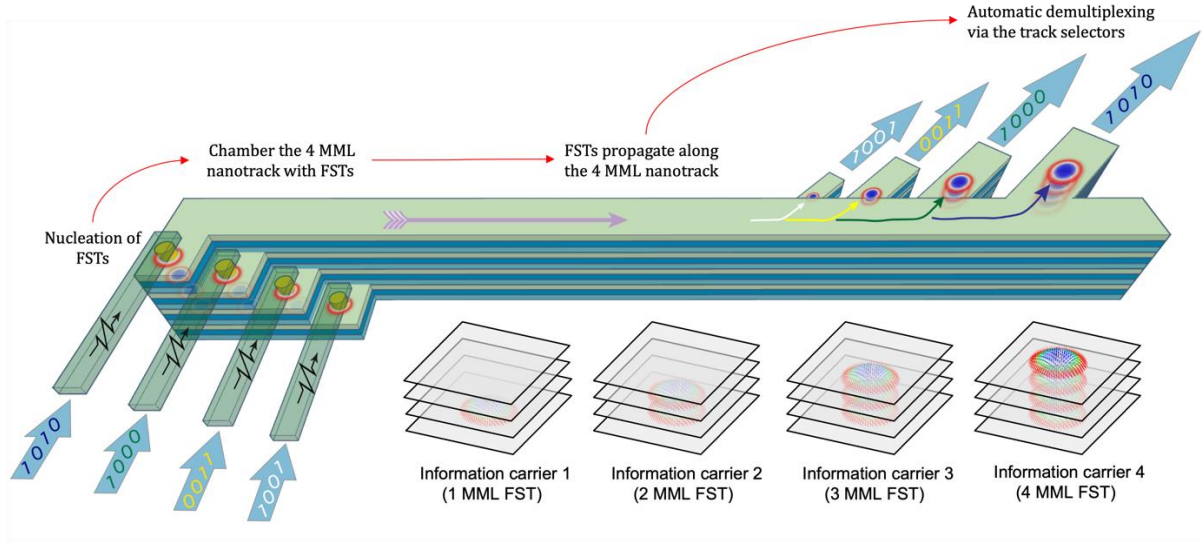

**Figure 3**. Schematic of the proposed use of FSTs in a multilayer nanotrack as a multiplexing device. The workflow of the proposed device is: nucleation of FSTs at the terrace-like MML stages, chambering the 4MML nanotrack with multiple FSTs, FSTs propagation along the 4MML nanotrack, and automatic demultiplexing of FSTs via the four-branched track selectors. A group of FSTs (i.e. 1MML FST, 2MML FST, 3MML FST, and 4MML FST) serve as information carries.

### 3.2. Nucleation of the fractional skyrmion tubes

The FSTs can be nucleated in the terrace-like MML stage structure shown in Fig. 4(a) in the proposed multilayer multiplexing device. There are individual MTJs on the surface of each step of the four-level stage, each of which is used to inject electric current providing STT for nucleating skyrmions in the MMLs beneath it. The main staircase structure and MTJs shown in Fig. 3 and Fig. 4(a) can be fabricated by additive or subtractive lithography processes. Single-layer precision may be achieved either through careful calibration of the material growth/removal rate or by dose-modulated electron-beam lithography [52]. In order to contact the top electrode of the MTJs for tunnelling in the perpendicular direction, each of them must be sheathed by an insulating material such as $SiO_2$.

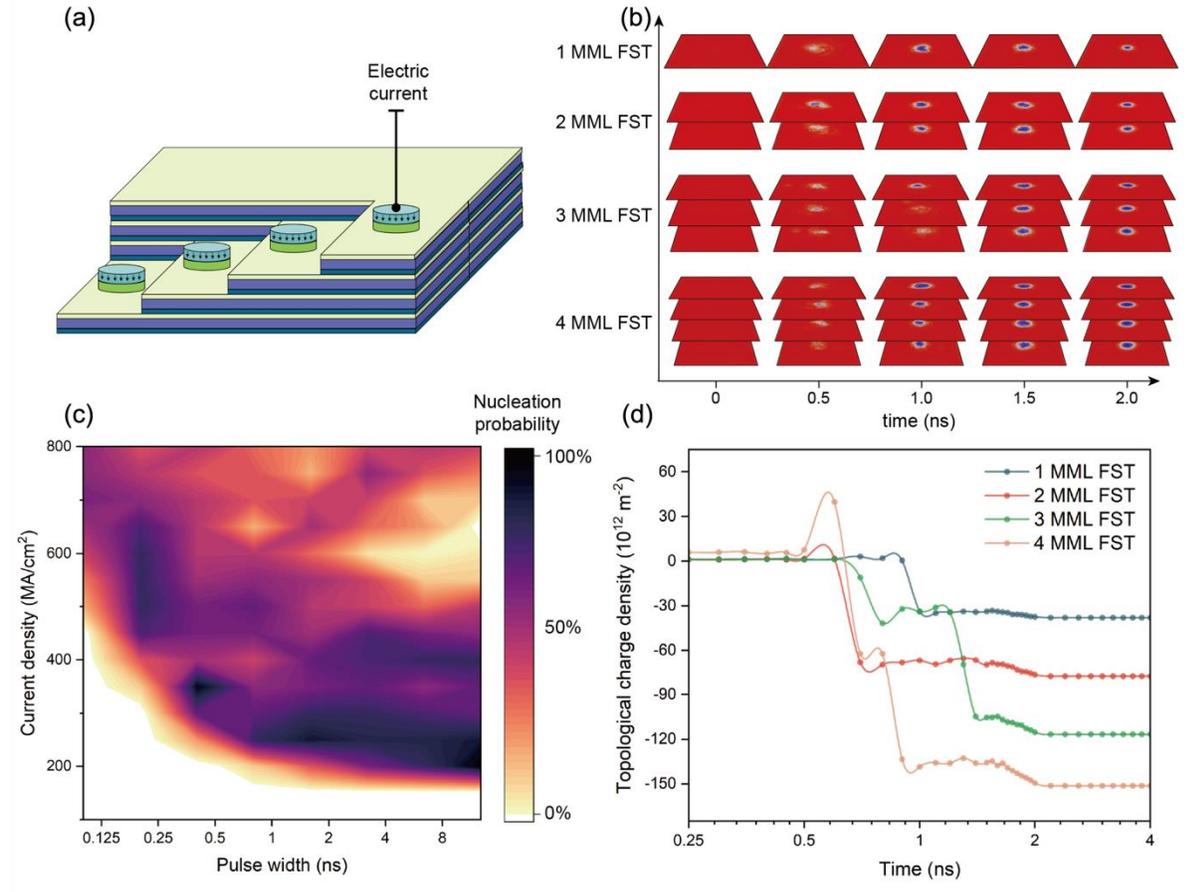

**Figure 4**. Nucleation of the fractional skyrmion tubes at a multilayer terrace-like stage in the proposed multiplexing device at room temperature. (a) Schematic drawing of the terrace-like nucleation site. An MTJ is placed on the surface of each terrace as an electric writing head. With this design, four FSTs can be nucleated individually. (b) The nucleation process of 1MML, 2MML, 3MML, and 4MML FSTs by a 1.5 ns width electric current pulse (full width at half maximum). The screenshots of each FM layer within FSTs at 0 ns, 0.5 ns, 1.0 ns, 1.5 ns, and 2.0 ns are presented in a timeline. (c) Probability phase diagram of successfully nucleating four FSTs

at room temperature, with varied pulse width (0.1 ns to 12.8 ns) and the amplitude of current density (100 MA cm$^{-2}$ to 800 MA cm$^{-2}$). Each data point illustrates the probability of successful nucleation of FSTs out of 20 distinct attempts. (d) Evolution of topological charge density for 1MML, 2MML, 3MML, and 4MML FSTs during the nucleation process in (b).

The simulated system has a four-level terrace geometry with cell size 2 nm x 2 nm x 1 nm. From left to right, the geometry of each terrace is 600 nm x 600 nm x 2nm, 600 nm x 600 nm x 4nm, 600 nm x 600 nm x 6nm, and 600 nm x 600 nm x 8nm. The MML has a small interlayer exchange coupling of $J_{\text{Interlayer}}$ = 0.02 mJ m$^{-2}$. The simulations were performed at room temperature by including an extra thermal fluctuation field [31] and with room temperature magnetic parameters. The magnetic parameters were retrieved from the literature of experimental measurements (see Methods). Electric current is injected into the MMLs from the MTJ using a CPP geometry, and the FSTs are nucleated through the STT. The pulse width of the injected current is 1.5 ns full-width at half maximum (FWHM) with a peak current density of 250 MA cm$^{-2}$. Although the pulse length is very short, the requisite current density is very large and may damage to the ultrathin tunnel barrier of the MTJ through Joule heating [53]. Therefore, future work is required to minimise the effects of Joule heating, either through optimisation of the spin injection process or by including a heatsink.

Fig. 4(b) presents each FM layer of the 1MML FST, 2MML FST, 3MML FST, and 4MML FST during the nucleation process. Each simulation is initialised in the uniform FM state at the beginning. After receiving the STT from the electric current pulse, the magnetisations in each case first experience a fluctuation period of around 0.5 ns, before individual skyrmions form in each MML. After a time frame of 1.5 ns, we can see from Fig. 4(b) that skyrmions are nucleated and stabilised in all four FSTs. To better understand the nucleation process of FSTs, the topological charge density for the four regions of each FST from left to right in Fig. 4(a) is extracted and displayed in Fig. 4(d). The topological charge density remains unchanged until 0.5 ns. From 0.5 ns to 1.5 ns, a few fluctuations in the topological charge density for FSTs can be seen before they reach the final skyrmion state. The final values of the topological charge density are approximately $40\times10^{12}$ m$^{-2}$, $80\times10^{12}$ m$^{-2}$, $120\times10^{12}$ m$^{-2}$, and $160\times10^{12}$ m$^{-2}$ for 1MML FST, 2MML FST, 3MML FST, and 4MML FST, respectively. The calculated topological charge density displayed in Fig. 4(d) verifies the results in Fig. 4(b).

Since a stochastic thermal field was included in the simulations of FST nucleation, the successful nucleation of FSTs occurs probabilistically, rather than being deterministic. We therefore determined the probability phase diagram of the FSTs by scanning the electric pulse width from 0.1 ns to 12.8 ns in a geometric sequence (i.e. 0.1, 0.2, 0.4, 0.8, 1.6, 3.2, 6.4, 12.8 ns) and the current density from 100 MA cm$^{-2}$ to 800 MA cm$^{-2}$ in 50 MA cm$^{-2}$ increments. We simulated the nucleation process for each FST with each given pulse width and current density 20 times, then calculated the probability of successful nucleation. The probability phase diagrams of successful nucleation of 1MML to 4MML FSTs are shown in Fig. S5 in Supporting Information. There are 15×8=120 data points displayed in the figures, where each data point illustrates the probability of successful nucleation out of 20 distinct attempts. The colour coding describes the probability value, where white represents 0% of the probability and black represents 100%. A transition from low nucleation probability for 1MML FST to high nucleation probability for 4MML FST can be seen in Fig. S5. However, for the multiplexing device proposed in this work, four different FSTs need to be nucleated at the four-stage terraces individually at the same time. Therefore, we also calculated the probability for successfully nucleating four FSTs simultaneously, with results summarised in the phase diagram displayed in Fig. 4(c). The results suggest that FSTs are more likely to be nucleated when we apply a large amplitude of current density (200 MA cm$^{-2}$ to 300 MA cm$^{-2}$) with a relatively large pulse duration (greater than 1 ns). This information is critical for experimental realisation of the multiplexer device. This information is critical for experimental realisation of the multiplexer device, wherein controlled generation of individual FSTs is required. There has been much recent progress in experimentally realising single-skyrmion nucleation in multilayer systems, including via defects generated by ion-irradiation [54], through local applied magnetic fields from a magnetic tip [55], and through spin-polarised currents [56].

We neglected the spin memory loss (SML) [46] when injecting STT to magnetisations in cascaded MMLs to simplify the nucleation procedure. Careful consideration of SML in the device-level simulations may be the focus of follow-up work. The results in this section demonstrate the nucleation of FSTs in the proposed multiplexing device, including the structure of the four-stage terrace nucleation site, the detailed nucleation process of FSTs, and the probability phase diagram of successful nucleation. In the following section, the second important stage of the device workflow will be addressed, namely the FST initialisation process.

### 3.3. Chambering the multiplexing device with FSTs

After nucleating FSTs in the four-stage terraces, the proposed device needs to be initialised before all of the FSTs propagate in the main 4MML nanotrack. The FSTs must propagate upwards from the nucleation sites into the main nanotrack during the initialisation process. We refer to the initialisation process as "chambering" in this work. FSTs must overcome an energy barrier when moving into the 4MML nanotrack from a film with a different number of MMLs [39], which can be attributed to the dipolar field in the two systems. As shown in Fig. 5(a), the 4MML multiplexing device needs to be chambered with all four FSTs from the four-stage terraces. We used CIP geometry for the chambering process because STT can act on every skyrmion throughout the FSTs, which decreases the risk of FSTs decoupling.

Fig. 5(a) illustrates the chambering process enabled by an STT resulting from the CIP in the $y$-direction with the current density of 100 MA cm$^{-2}$ and pulse width of 4 ns. FSTs first propagate towards the boundary of different MMLs systems in the $y$-direction, i.e. along the direction of the applied current, before moving along the boundary in the $x$-direction for a short distance. They subsequently cross the boundary and successfully move into the main track. The 1MML FST exhibits the largest displacement along the $x$ direction; this displacement gradually reduces as the number of MML repeats increases. We then varied the interlayer exchange coupling constant $J_{Interlayer}$ and the amplitude of the current density to generate a phase diagram of the initialisation process. By changing $J_{Interlayer}$ and current density, the initial state leads to four phases which we label "chamber", "stuck", "nucleate" and "decouple", as shown in the left panel of Fig. 5(b). If the interface traps the FST, we mark it as a "stuck"; if all four FSTs propagate into the main track, we mark it as a "chamber"; if any of FSTs annihilate, we mark it as a "decouple"; if the skyrmions in different layers separates we also mark this as a "decouple"; if a new skyrmion is nucleated within the FM layers during the process, we mark it as a "nucleate".

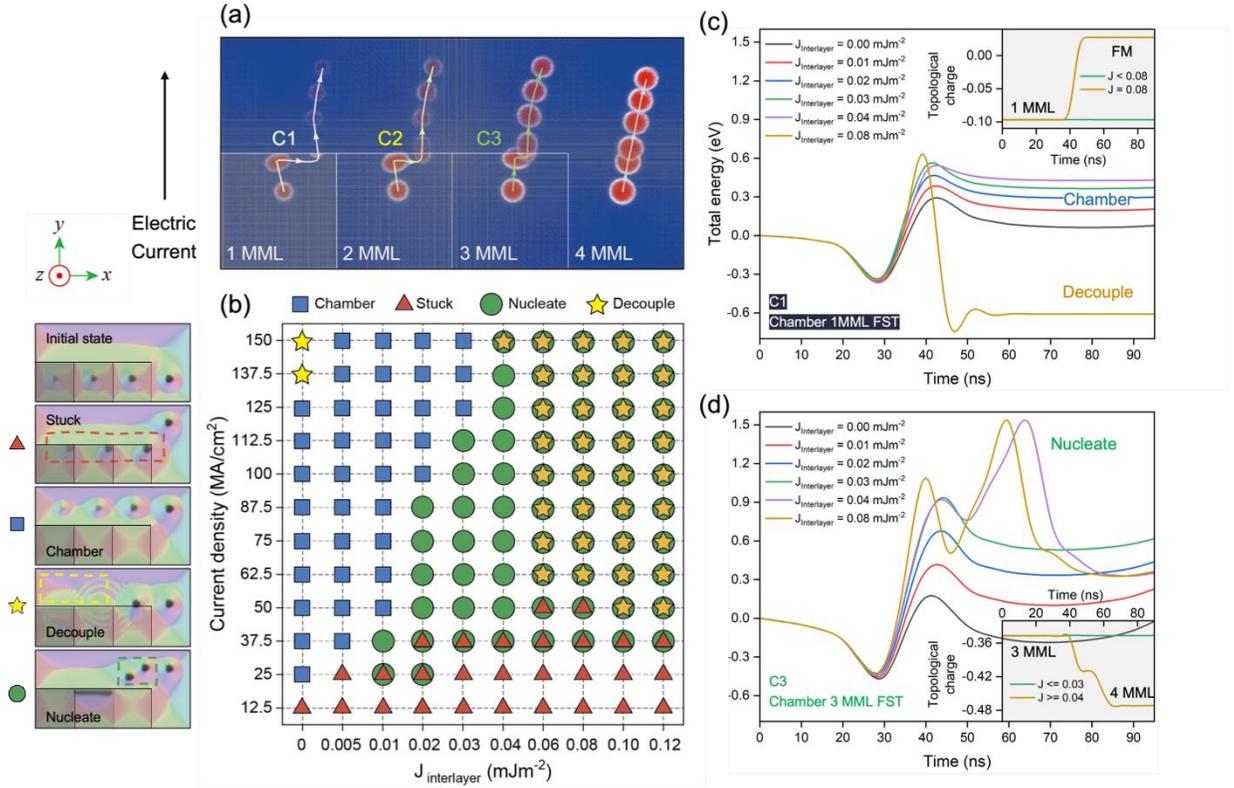

**Figure 5**. Chambering the multiplexing device with FSTs (the initialisation process). (a) Micromagnetic simulation results of chambering the 4MML nanotrack with a 1MML FST, a 2MML FST, a 3MML FST, and a 4MML FST. The trajectories of each FST are marked as C1, C2, and C3, respectively. The magnetic parameters used were: interlayer exchange coupling constant $J_{Interlayer}$ = 0.02 mJ m$^{-2}$, external magnetic field 90 mT, and DMI constant 1.9 mJ m$^{-2}$. (b) Phase diagram of chambering four FSTs into the main track. The panel to the left of the phase diagram shows examples from the simulations of each of the behaviours. The red dashed rectangle highlights the case that some or all of the FST states get stuck at the interface between the terraces and the main nanotrack. The yellow dashed rectangle demonstrates the case of FST decoupling, where one or more of the FSTs annihilate. Energy paths for the chambering process C1 and C3 are shown in (c) and (d), respectively. The energy path for process C2 is shown in Fig. S6 in Supporting Information.

The results of the initialisation process after injecting electric current are summarised in the phase diagram shown in Fig. 5(b). The four possible phases "chamber", "stuck", "nucleate", and "decouple" are marked with a blue rectangle, a red triangular, a green circle, and a yellow star respectively. As the target phase, the "chamber" phase has a decently large parameter window, which occupies the upper left 1/4 of the phase diagram. In the multilayers with large interlayer exchange coupling constants ($J_{\text{Interlayer}} \geq 0.06$ mJ m$^{-2}$), 1MML FSTs are stuck by the boundary when applying insufficient current densities ($J_{\text{dc}} < 50$ MA cm$^{-2}$) and will annihilate when the current density is large ($J_{\text{dc}} \geq 50$ MA cm$^{-2}$). 3MML FSTs will nucleate a new skyrmion in the top FM layer when crossing the boundary with high interlayer exchange coupling constants ($J_{\text{Interlayer}} > 0.03$ mJ m$^{-2}$) and applied current densities ($J_{\text{dc}} > 37.5$ MA cm$^{-2}$). 2MML FSTs behave more robustly and produces the "chamber" phase in most situations except for at extremely high interlayer exchange coupling constants ($J_{\text{Interlayer}} > 0.08$ mJ m$^{-2}$) and current densities ($J_{\text{dc}} > 125$ MA cm$^{-2}$), wherein annihilation is observed. 4MML FSTs always produce the "chamber" phase, resulting from the fact that there is no energy barrier during the propagation process in this case. The superposed markings in the phase diagram indicate different behaviours from different FST types. For instance, all cases of the yellow stars superposing green circles correspond to the situation that the 1MML FST annihilates and the 3MML FST nucleates a new skyrmion; the red triangle with the green circle indicates that the 1MML FST is stuck and the 3MML FST nucleates a new skyrmion.

In order to better explain the phases shown in Fig. 5(b), we extracted the total micromagnetic energy of the regions containing 1MML FST, 2MML FST, 3MML FST, and 4MML FST individually. The trajectories when chambering 1MML FST, 2MML FST, and 3MML FST are marked as C1, C2, and C3, respectively, in Fig. 5(a). The energy evolution of C1, C2, and C3 can be seen from Fig. 5(c), Fig. S6 of Supporting Information, and Fig. 5(d). Here we varied the interlayer exchange coupling constant from 0 to 0.08 mJ m$^{-2}$ while fixing the current density $J_{dc}$ = 125 MA cm$^{-2}$. With a higher interlayer exchange coupling constant, the "decouple" phase is observed in C1, while the "nucleate" phase could be seen in C3. The energy evolution is further demonstrated by the change of topological charge of the system shown in the insets of Figs. 5(c) and 5(d). The energy barrier for chambering 1MML FST into the central track increases as a linear relationship with the amplitude of $J_{Interlayer}$. When chambering 1MML FST, the "chamber" phase is obtained when $J_{Interlayer} \leq 0.06$ mJ m$^{-2}$; therefore, no net change of topological charge is observed. Annihilation of the skyrmion happens at $J_{Interlayer} = 0.08$ mJ m$^{-2}$, so the topological charge of the system changes to zero. Similarly, when chambering the 3MML FST, a net increase of the topological charge by around 1/3 occurs, suggesting the nucleation of a new skyrmion in the system at higher interlayer exchange coupling constants ($J_{Interlayer} \geq 0.04$ mJ m$^{-2}$). Note that the initialisation process of 2MML FST, marked as C2 in Fig. 5(a), merely exhibits "chamber" and "stuck" phases in Fig. 5(b), and therefore no noticeable variation can be observed in the topological charge. The energy evolution when chambering the 2MML FST is presented in Fig. S6 of the Supporting Information.

### 3.4. Voltage-controlled synchronizers for pipelining

After the initialisation process, we now have the multiplexing device chambered with FSTs in the main nanotrack. FSTs can then be transmitted along the track towards the demultiplexing region. Considering the long transmission distances likely required in real-world applications, it is better to divide the transmission track into several regions, as shown in Fig. 6(a). In such a design, we can achieve pipelined transmission for information carriers to enhance device throughput. The results in Fig. 2 have demonstrated that FSTs with different MMLs propagate with various velocities, where the 4MML FST moves the most slowly, and the 1MML FST the most quickly. Therefore, synchronizers are required to maintain the correct order of information sequences and avoid sticky packets during data transmission [25]. Here we can utilise voltage gates as synchronizers based on the voltage-controlled-magnetic-anisotropy (VCMA) effect. The VCMA effect was first reported in a 3D transition ferromagnetic layer in 2–4 nm thick FePt (Pd) films [57]. Surprisingly, it has been reported that a small electric field of 100 mV nm$^{-1}$ is sufficient to change the PMA by 40%, corresponding to a VCMA efficiency of 210 fJ V$^{-1}$ m$^{-1}$ at room temperature [58]. In this work, the simulation of the VCMA effect is based on a linear relationship [12]:

$$K_{uv} = K_{u0} + \vartheta V_b, \tag{7}$$

where $\vartheta$ is the VCMA coefficient, $V_b$ is the bias voltage on the VCMA gate, and $K_{u0}$ is the background anisotropy constant. Regions in which the PMA is elevated provide an increased energy barrier, whereas regions of reduced PMA lead to a potential well. When simulating the pipelined transmission of FSTs schematised in Fig. 6(a), we periodically set VCMA gates on and off by changing the amplitude of the PMA constant by 10%. An electric driving current with the amplitude of 3 MA cm$^{-2}$ is applied in the HM layer in the *x*-direction. Compared to the chambering process of FSTs where STT is utilized, we apply SOT for the transmission of FSTs instead to achieve higher driving efficiency and thus lower energy consumption [13].

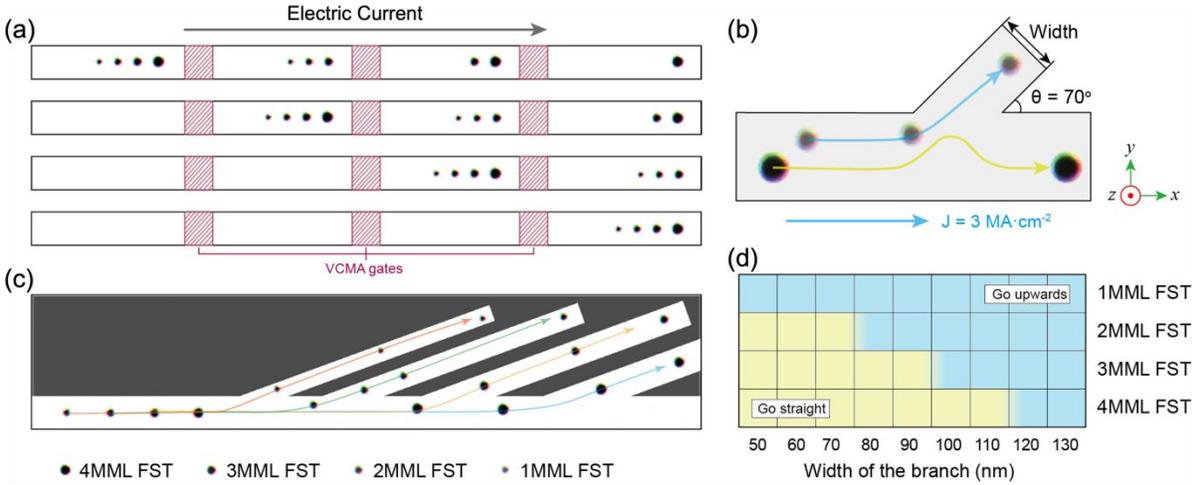

**Figure 6**. Pipelined transmission and automatic demultiplexing of FSTs in the proposed device. (a) Schematic showing pipelined transmission of four FST packets. Each information packet consists of a combination of multiple FSTs. The red shaded areas denote the VCMA synchronizers. (b) Schematic illustration for tuning the branch width of the track selector. SOT drives the FSTs with the current density of 3 MA cm$^{-2}$. (c) Simulation results of automatically demultiplexing FSTs via a four-branch track selector. (d) Track selecting phase diagram for 1MML FST, 2MML FST, 3 MML FST, and 4MML FST, respectively, while varying the width of the branch and fixing the current density at 3 MA cm$^{-2}$.

The pipelined transmission of four FSTs in the 4MML main track is illustrated in Fig. 6(a). We deployed three VCMA gates in the track, which are marked as red shaded rectangles. The track is divided into four regions as registers, each of which can store a packet of information consisting of four bits. The presence (absence) of a 1MML FST/2MML FST/3MML FST/4MML FST encodes a single binary "1" ("0") for the first/second/third/fourth bit respectively. Initially, four packets of information signals "1111", "1110", "0011", and "0001" are positioned in consecutive registers of the device via propagation mediated by the VCMA gates. They are then transmitted along the *x*-direction and synchronised by adjacent VCMA gates. It should be noted that the FSTs are nucleated and propagated from left to right in order of the number of MML repeats, such that within an information packet, the order is [1MML FST | 2MML FST | 3MML FST | 4 MML FST]. This design can retain the order of FSTs and prevent mismatch between information packets during transmission. Fig. 6(a) indicates that the information integrity and sequence order are well retained, allowing the transmission efficiency to be quadrupled. Note that for clarity, Fig. 6(a) is a schematic representation of the VCMA-controlled FST transmission. The full simulation results are provided in Video S1 of the Supporting information. Furthermore, this pipelined design is not limited to four-bit packets of information. The VCMA gates can support as many information carries in the nanotrack as can be stabilised in the MML system. Here we use four carries in each packet for illustration and consistency with the other results in this paper.

### 3.5. Automatic demultiplexing via track selectors

The final stage of the proposed MML multiplexing device workflow is to filter the FSTs out of the information packets to decode the information signals. Here we propose to use a four-branch track selector to filter the FSTs into individual tracks for detection. We can utilise a track selector to filter different FSTs, as illustrated in Fig. 6(b). As has been discussed above, there are two choices to filter the FSTs: the first one is via the skyrmion Hall angle $\theta_{SkHE}$ and the second is the size difference of the skyrmion diameter. Changing the angle or the width of the branch can filter and demultiplex different FSTs. As for the skyrmion Hall angle $\theta_{SkHE}$, our results suggest that there is a difference in $\theta_{SkHE}$ for FSTs with different MMLs. Namely, $\theta_{SkHE}$ for 1MML FST to 4MML FST under SOT is -47.5°, -44.0°, -40.7°, and -35.0°, respectively. At the same time, there is also a difference in the skyrmion diameter among FSTs, where the skyrmion diameter for 1MML FST to 4MML FST is 21.5 nm, 26.5 nm, 32.5 nm, and 38.0 nm, respectively. The angle-based selector would require a much larger device footprint to filter the FSTs for their relatively small angle difference. Therefore, in this work, we propose to demultiplex the information signals by a width-based track selector.

Simulations of the demultiplexing process were performed to obtain a phase diagram for the four-branch track selector. Here, we modified the width of the branch from 50 nm to 130 nm while the angle of the branch was fixed at 70°, and the electric current density was 3 MA cm$^{-2}$ (see Fig. 6(b)). The four FSTs individually propagated through the selector along the track, driven by the SOT. If the FST propagated into the branch, we marked the result as "go upwards". Otherwise, it was marked as "go straight". The track selecting phase diagram for 1MML to 4MML FSTs is shown in Fig. 6(d), where the sky-blue coloured region represents FSTs propagating into the branch and the cream-yellow coloured region denotes FSTs moving straight, passing by the branch. The results indicate enough space to design a track selector for filtering four FSTs out of one packet because each FST has an exclusive window with 20 nm width. Guided by the track selecting phase diagram for FSTs, we configured a four-branch track selector as shown in Fig. 6(c). The width of these four branches is 60 nm, 80 nm, 100 nm, and 120 nm for demultiplexing 1MML FSTs, 2MML FSTs, 3MML FSTs, and 4MML FSTs respectively. Successful demultiplexing of the FSTs shown in Fig. 6(c) verifies the feasibility and validity of such a width-based track selector.

In summary, these results demonstrate that the proposed fractional skyrmion tubes in the MML system can be potential candidates as information carriers, based on the fact that multiple FSTs can be nucleated, transmitted, and filtered in a single MML device. The results highlight the potential for three-dimensional skyrmionics quasiparticles to significantly expand the information storage and processing capability of tailored magnetic multilayer systems.

**CONCLUSIONS**

This study was designed to assess the hypothesis that in MML systems, skyrmions can exist within part of a multilayer, as fractional skyrmion tubes. We confirmed this with magnetic energy analysis and micromagnetic simulations. The findings suggest that distinct FST states may coexist in a single MML system which are tuneable in their thermal stability and magnetic properties. Their topological properties and current-driven behaviour were analysed both with theoretical calculations and micromagnetic simulations. This work also proposes to use such FSTs to encode information in an MML multiplexing device, where multiple FSTs can be nucleated, transmitted, and filtered. This study provides the first comprehensive assessment of encoding information by multiple FSTs in a single MML device, highlighting the potential utility of distinct skyrmion states in magnetic multilayer systems. Further investigations are required to explore device settings for FSTs and establish effective nucleation and detection methods.

**METHODS**

**Micromagnetic simulations:**

The micromagnetic simulations were performed using the GPU-accelerated micromagnetic programme Mumax$^3$ [31]. The time-dependent magnetisation dynamics are conducted by the Landau-Lifshitz-Gilbert (LLG) equation:

$$\frac{d\mathbf{m}}{dt} = -|\gamma_{\text{LL}}|\mathbf{m} \times \mathbf{h}_{\text{eff}} + \alpha \mathbf{m} \times \frac{d\mathbf{m}}{dt} + \mathcal{T}_{SOT}\mathbf{m} \times (\mathbf{m}_{\text{p}} \times \mathbf{m}) \qquad (8)$$

where $\mathbf{m} = \mathbf{M}/M_s$ is the reduced magnetisation, $M_s$ is the saturation magnetisation, $\gamma_{LL}$ is the gyromagnetic ratio, $\mathbf{h}_{eff} = \mathbf{H}_{eff}/M_s$ is the reduced effective field, $\alpha$ is the damping parameter, $\mathcal{T}_{SOT} = \frac{\gamma_e \hbar j_e \theta_{SH}}{2 e M_s t}$ is the SOT efficiency with $\gamma_e = \frac{\gamma}{\mu_0} = 1.76 \times 10^{11}$ T$^{-1}$ s$^{-1}$ being the gyromagnetic ratio of an electron, $\hbar$ is the reduced Planck constant, $j_e$ is the current density, $\theta_{SH}$ is the spin Hall angle, $e$ is the electron charge, $t$ is the thickness of the FM layer, and $\mathbf{m}_p$ is the polarization direction of the spin current. The energy density $E$ is a function of $\mathbf{m}$, which contains the exchange energy term, the anisotropy energy term, the Zeeman energy term, the magnetostatic energy term, and the DMI energy term. The material parameters to perform the simulations are chosen according to previous reported room temperature experimental results [20]: damping parameter $\alpha = 0.1$, DMI constant $D_{int} = 1.5$ mJ m$^{-2}$ to 2.6 mJ m$^{-2}$, the value for Gilbert gyromagnetic ratio $\gamma = -2.211 \times 10^5$ mA$^{-1}$ s$^{-1}$, saturation magnetization $M_s = 956$ kA m$^{-1}$, the spin Hall polarisation $\Theta_{SH} = 0.6$ to enhance the spin Hall effect, the uniaxial out-of-plane magnetic anisotropy $K_u = 717$ kJ m$^{-3}$, the polarisation of the spin current is in the $+y$ direction, and the exchange constant is assumed to be A = 10 pJ m$^{-1}$. To ensure the accuracy of calculation, the mesh size is set to 2 nm × 2 nm × 1 nm, which is smaller than the exchange length $l_{EX} = 2\sqrt{A/(\mu_0 M_s^2)} = 6.0$ nm and DMI length $l_{DMI} = \frac{2A}{D_{int}} = 11.8$ nm. An external magnetic field of 10 mT to 100 mT in the out-of-plane direction is applied for the simulations. In the simulated MMLs, the intermediate HM$_1$ and HM$_2$ layers are thinner than the electron spin diffusion length. In this case, the torques would be efficient only in the external layers. In the simulation of FSTs propagation along the nanotrack, the SOT created via a CPP is applied only in the bottom layer, and the injected spin polarization is uniform in the layer. The injected current is then modelled as a fully polarized vertical spin current. In the simulation of the chambering process, electric current is applied in the $+y$ direction via the CIP geometry, where skyrmions in all FM layers within FSTs are driven by STT.

**ASSOCIATED CONTENT**

**Supporting Information**

Figures S1-S6 of the supporting information PDF show:

(S1) Phase diagrams for FST stability with respect to DMI and external magnetic field in a four-layer system for different values of the interlayer exchange constant;

(S2) as in Fig. S1, but for an eight-layer system;

(S3) the binding energy of different FST states for four values of the interlayer exchange constant;

(S4) the binding energy over a range of interlayer exchange couplings for 2MML, 3MML, 4MML, and 6MML FSTs;

(S5) a phase diagram of the nucleation probability for different FST states as a function of applied current density and pulse width, and;

(S6) the evolution of the total micromagnetic energy during the chambering process for a 4MML multiplexing device.

Supporting Video S1 shows a simulation of VCMA-controlled FST propagation.

**AUTHOR INFORMATION**

**Corresponding Author**

christoforos.moutafis@manchester.ac.uk

**Author Contributions**

R. C. and C. M. conceived the project, and R. C., C. M., Y. L., and Y. Z. contributed to the project design. R. C. and Y. L. performed the micromagnetic simulations and theoretical calculations. R. C., C. M., and W. G. prepared the manuscript. All authors discussed and commented on the manuscript. All authors have approved the final version of the manuscript.

**ACKNOWLEDGMENT**

This work is supported by the Engineering and Physical Sciences Research Council (EPSRC) under the grant 'Skyrmionics for Neuromorphic Technologies', EP/V028189/1. The authors would also like to acknowledge the assistance provided by Research IT and the use of the Computational Shared Facility at the University of Manchester. R. C. and Y. L. wish to acknowledge the Department of Computer Science Kilburn Scholarship for the funding support. R. C. would like to thank the University of Manchester President's Scholarship. The authors would like to acknowledge discussions on the manuscript with Ioannis Charalampidis.